\documentclass[article, shortnames, nojss]{jss}
\usepackage{minted}
\author{Olga Lezhnina\\TIB Hannover \And 
        Manuel Prinz\\TIB Hannover \And
        Markus Stocker\\TIB Hannover}
\title{\pkg{dtreg}: Describing Data Analysis\\ in Machine-Readable Format in Python and R}

\Plainauthor{Olga Lezhnina, Manuel Prinz, Markus Stocker}
\Plaintitle{dtreg: Describing Data Analysis in Machine-Readable Format in Python and R} 
\Shorttitle{\pkg{dtreg}: Machine-Readable Description of Data Analysis} 

\Abstract{
  For scientific knowledge to be findable, accessible, interoperable, and reusable, it needs to be machine-readable. Moving forward from post-publication extraction of knowledge, we adopted a pre-publication approach to write research findings in a machine-readable format at early stages of data analysis. For this purpose, we developed the package \pkg{dtreg} in Python and R. Registered and persistently identified data types, aka schemata, which \pkg{dtreg} applies to describe data analysis in a machine-readable format, cover the most widely used statistical tests and machine learning methods. The package supports (i) downloading a relevant schema as a mutable instance of a Python or R class, (ii) populating the instance object with metadata about data analysis, and (iii) converting the object into a lightweight Linked Data format. This paper outlines the background of our approach, explains the code architecture, and illustrates the functionality of \pkg{dtreg} with a machine-readable description of a t-test on Iris Data. We suggest that the \pkg{dtreg} package can enhance the methodological repertoire of researchers aiming to adhere to the FAIR principles.
}

\Address{
  Olga Lezhnina\\
  TIB – Leibniz Information Centre for Science and Technology\\
  30167 Hannover, Germany\\
  E-mail: \email{Olga.Lezhnina@tib.eu}   
  
  \vspace{5mm}
  
  Manuel Prinz\\
  TIB – Leibniz Information Centre for Science and Technology\\
  30167 Hannover, Germany\\
  E-mail: \email{Manuel.Prinz@tib.eu} 
  
  \vspace{5mm}

  Markus Stocker (the corresponding author)\\
  TIB – Leibniz Information Centre for Science and Technology\\
  30167 Hannover, Germany\\
  E-mail: \email{Markus.Stocker@tib.eu} 
}

\begin{document}

\section[Introduction]{Introduction}

Scientific knowledge should be findable, accessible, interoperable, and reusable (FAIR), which means that it should be, among other properties, machine readable \citep{wilkinson2016fair}. Extracting machine-readable knowledge from already published research papers is conducted either manually by human experts or via automated methods; the former approach is time-consuming, while the latter is prone to errors, and involving large language models (LLMs) is far from being a panacea \citep{huang2025survey, openai2025}. Therefore, some researchers are also looking into pre-publication approaches, which ensure machine readability and accurate description of data and processes at early stages of the research life cycle.   

We developed the package \pkg{dtreg} in Python \citep{python} and R \citep{rcore} in the frame of a pre-publication approach aimed at structuring research findings in accordance with registered and persistently identified  data types, aka schemata, and writing them in a machine-readable format \citep{stocker2025rethinking, 10.1145/3726302.3730134}. Currently, \pkg{dtreg} supports data types implemented in the European Persistent Identifier Consortium (ePIC, \url{https://pidconsortium.net}) and the Open Research Knowledge Graph (ORKG, \url{https://orkg.org}). The use of \pkg{dtreg} is fairly easy: first, a mutable instance of a schema-related Python or R class is created. After the researcher populates the instance with data analysis information, it is written into the machine-readable format JavaScript Object Notation for Linked Data (JSON-LD). This format supports references to identifiers on the web, is lightweight, readable for humans and machines, and therefore widely used in FAIR data management \citep{sporny2014json, musen2022modeling}. 

Our goal is to empower researchers to make their findings machine readable with easy-to-use practices giving them more control over the process while not interfering with their established procedure of data analysis. In the rest of the paper, we explain the role of data analysis schemata, outline related work, give insight into the code architecture, and show how to use \pkg{dtreg} to write research findings in JSON-LD  format and meet an important aspect of FAIR data. We did not use any AI technologies in developing \pkg{dtreg} and writing this paper. 

\section{Background}

\subsection{Data type registries (DTRs)}\label{2.1}
The name of the \pkg{dtreg} package stems from data type registries (DTRs), as it is designed to make research findings machine readable, and for that, these findings are structured in accordance with data types from a DTR. Data types are identified, defined, and registered characterizations of data at any level of granularity \citep{broeder2014data, lannom2015rda, ma2016semantic}. DTRs describe data types and assign persistent identifiers to these types for resolving them in an unambiguous way. 

Currently, the \pkg{dtreg} package supports two DTRs, the ePIC type registry\footnote{See \url{https://typeregistry.pidconsortium.net}.} and the ORKG templates. The ePIC is an identifier system for the registration, storing, and resolving of persistent identifiers for scientific data \citep{kalman2012european}. The ePIC consortium includes research infrastructures and data centers from different countries and is open to any organization that stores research data. The ORKG is an infrastructure for the production, curation, publication, and use of FAIR scientific knowledge \citep{stocker2023fair}. The ORKG initiative actively engages research communities and intergovernmental organizations all over the world \citep{orkg2024} and integrates crowdsourcing with automated data extraction techniques for creating scholarly knowledge graphs \citep{verma2023scholarly}.   

We constructed a number of data types, aka schemata\footnote{We prefer the term "schemata" for convenience reasons, but they are, strictly speaking, registered and persistently identified data~types.}, in the frame of the ePIC DTR to structure methods, data, and results of data analysis. They are described in detail in the next section.

\subsection{Data analysis schemata}\label{2.2}
Data analysis schemata assist researchers in specifying their findings in a structured manner. The current version of schemata comprises data analysis methods most widely used in computer science, environmental sciences, and other domains. 

Categorization of data analysis methods is a complicated task that can be viewed from different perspectives \citep{vorberg1999auswahl, ranganathan2021introduction}. We adhered to commonly accepted categorizations, e.g., descriptive versus inferential statistics, regression versus correlation versus comparing group means, categorical versus interval target/dependent variables \citep{field2012discovering}, and  multilevel versus non-multilevel methods \citep{harrison2018brief}. As integration of statistical and machine learning methods has been discussed for decades \citep{breiman2001statistical, trevor2009elements}, we did not make a distinction between these areas; in many cases, such as regression or clustering, this distinction would not be possible, indeed. We also relied on widely used ontologies as presented in the Ontology Lookup Service \citep{jupp2015new}. The ontologies used to create the schemata include the Basic Formal Ontology \citep{Otte2022-OTTBBF}, the Information Artifact Ontology \citep{ceusters2012information}, the Semanticscience Integrated Ontology \citep{dumontier2014semanticscience}, the Statistical Methods Ontology \citep{lloyd2020struct}, and the Software Ontology \citep{malone2014software}.

\begin{figure}
    \centering
    \includegraphics[width=1\linewidth]{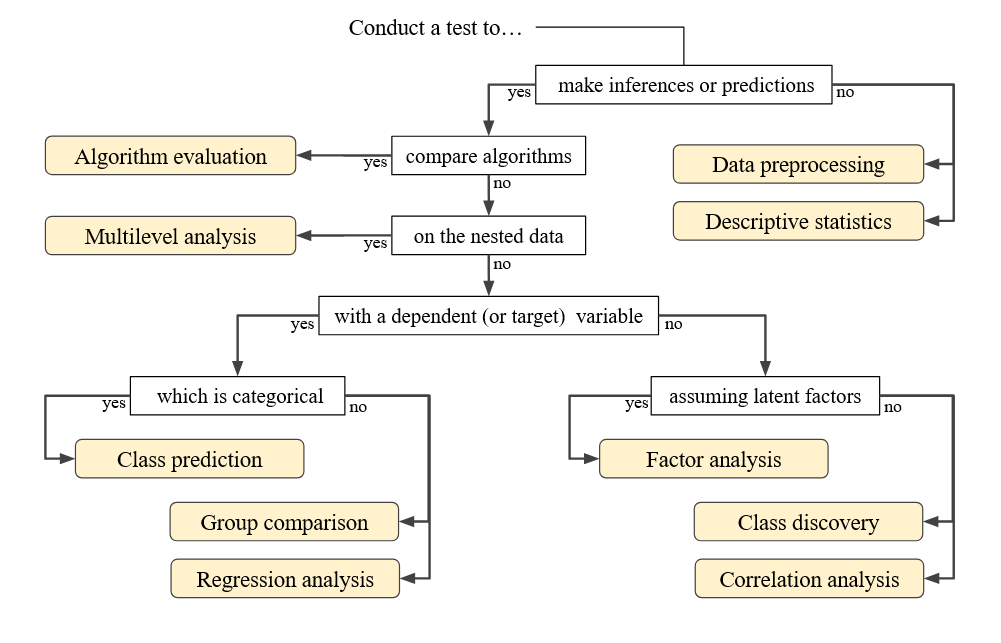}
    \caption{Selecting a schema for a data analysis method. Schemata are shown in yellow boxes, analytical choices in white boxes.}
    \label{fig_1}
\end{figure}

Selecting a schema to report a statistical test or a machine learning method is straightforward (see Figure~\ref{fig_1}). The names of schemata are self-explanatory. The \code{algorithm_evaluation} schema refers to a benchmark-based model evaluation, \code{multilevel_analysis} to hierarchical/mixed/nested models, and \code{group_comparison} to any comparison of two or more means within or between groups, such as t-tests, any ANOVAs, and their nonparametric analogues. The \code{class_discovery} schema describes clustering, and \code{class_prediction} any classification task in machine learning or logistic/ordinal regression. For more information, the user is referred to our help page  at \url{https://knowledgeloom.tib.eu/pages/help}, which also gives the URLs of these schemata.

Each of these data analytic schemata includes reusable sub schemata specifying the software method, the input data, and the output data. They are organized hierarchically: an analytic schema, such as \code{group_comparison}, includes \code{data_item} for both input and output, which in turn includes \code{table}, etc. Finally, the analysis is written in the overarching \code{data_analysis} schema and linked to a scientific statement \citep{hars2013publishing} to present research findings in a natural language.

\subsection{The Loom approach to machine-readable knowledge production}
 
We developed the package \pkg{dtreg} to support a pre-publication approach to machine-readable knowledge production \citep{stocker2025rethinking} in the frame of the TIB Knowledge Loom, an emerging open science digital library for analysis-ready scientific knowledge. The Loom approach aims at facilitating reuse, synthesis, integration, and transfer of scientific knowledge in accordance with the FAIR principles. It shares these goals with the ORKG but involves different methods to achieve them \citep{10.1145/3726302.3730134}. The core of the approach is to produce scientific knowledge in a machine-readable format at early stages of data analysis. When research findings are obtained in a computing environment, they can be structured in accordance with registered data types, and the resulting Python or R object can be converted into JSON-LD format. Here, we briefly outline recent amendments to the approach, in particular those related to \pkg{dtreg}.

Previously, the Loom approach relied on the existing \pkg{orkg} Python package \citep{orkg_py} and the code that we developed with the same functionality but different architecture in R. At that stage, the approach was restricted to ORKG-specific schemata (i.e., ORKG templates), and API requests were always needed to load the schemata. We addressed these limitations by developing the new schemata (as discussed in section \ref{2.2}) and implementing them in the ePIC DTR to ensure the reliable governance, which is an advantage over the crowdsourced ORKG templates. With the new schemata, we made a step towards improving the semantic interoperability of the data by relying as much as possible on terms from widely used ontologies, although full OWL-based formal semantics is yet to be achieved. Our new package \pkg{dtreg} supports the ePIC and ORKG DTRs, and any other DTR can be added by request, as allowed by functionality of the package (see section \ref{4.2}). The code was developed with the same architecture in Python and R, so that simultaneous change is easily implemented whenever required. Including the schemata in the \pkg{dtreg} static files supports writing data analysis results in a machine-readable format without API requests (which is done automatically if a requested schema is not available as a  static file); thus, the process becomes not only faster but also independent from any possible issues with DTR APIs or internet connectivity.

\section{Related work}

The aim of \pkg{dtreg} is to assist the user in writing machine-readable research findings and meet important criteria of the FAIR data principles for scientific knowledge. With the related purpose, a number of systems were developed that  share computational workflows and trace the provenance of scientific outputs to support FAIR data \citep{wilkinson2025applying}. These include Kepler \citep{ludascher2006scientific}, Common Workflow Language \citep{10.1145/3486897}, Fair Data Pipeline for epidemiological modeling \citep{mitchell2022fair}, Galaxy for biomedical research \citep{galaxy2024galaxy}, Reproducible Research Publication Workflow \citep{peer2022reproducible}, Apache Taverna or Taverna Workbench \citep{4534278}, and WorkflowHub \citep{gustafsson2025workflowhub}. Typically, such workflow management systems handle complex data processing workflows, deal with resource allocation and distributed task execution, and involve cloud resources \citep{wilkinson2025applying}. In comparison, \pkg{dtreg} focuses on data typing, and could be integrated for this task in a workflow management system.  

As \pkg{dtreg} helps to structure data analysis results in accordance with schemata, it can be compared to previous work on standardization required for machine-readable data. For example, the Cooperation Databank collected studies on human cooperation and developed an ontology to structure the results in a standardized format \citep{spadaro2022cooperation}. Formalization of reporting guidelines in life sciences and development of metadata schemata with the use of \code{schema.org}, a widely accepted DTR, was the focus of work by \cite{batista2022machine}. These different approaches, as well as ours, strive for syntactic and semantic interoperability and machine-actionable scientific data. 

Finally, and more specifically related to \pkg{dtreg} as a package, there are Python and R packages designed for research data management aimed at making research data FAIR. In Python, \pkg{PyRDM} assists in automated online publication of scientific software with input and output data \citep{jacobs2014pyrdm}. In R, the package \pkg{archivist} can be used to retrieve and validate R objects and increase research reproducibility \citep{JSSv082i11}. Also in R, the package \pkg{scienceverse} automatically evaluates predictions made in pre-registered studies by comparing them to the actual data provided by the researcher and reports the results in a machine-readable format, a workflow that is useful for conducting meta-analyses \citep{lakens2021improving}.  Rather than automating data publishing in general or evaluating research hypotheses, \pkg{dtreg} focuses on structuring research findings and data representation in a machine-readable format.  

\section{Functionality}

\subsection[Documentation and code quality]{Documentation and code quality}
The \pkg{dtreg} package in Python version \citep{dtreg_py} and R version \citep{dtreg_r} was released in PyPi and the Comprehensive R Archive Network (CRAN) respectively. The code is open access under the MIT license. The current version is v1.1.2, and changes from the previous versions are specified in the respective changelogs. In Python, we follow PEP8 style guidelines \citep{van2001pep}, and in R, the conventions for package development \citep{wickham2019advanced, wickham2023r}. In addition to documenting \pkg{dtreg} via \code{README} and API documentation, we included the vignette \textit{Introduction to dtreg} in the R version of the package to give users detailed explanation of its functionality. The R version of dtreg passed the CRAN checks, with results accessible at \url{https://cran.r-project.org/web/checks/check_results_dtreg.html}. We continuously monitor the test coverage through test reports generated by a GitLab build pipeline. Reports are generated by the \pkg{coverage} package in Python \citep{coverage_py} and the \pkg{covr} package in  R \citep{covr}. Currently, 95 percent of the code in  Python and 96 percent in  R are covered by unit tests. We support feedback from users with issue trackers in GitLab\footnote{For dtreg-Python, see \url{https://gitlab.com/TIBHannover/lki/knowledge-loom/dtreg-python/-/issues}, and for dtreg-R \url{https://gitlab.com/TIBHannover/lki/knowledge-loom/dtreg-r/-/issues}}, and further inquiries can be sent to our contact email \url{knowledgeloom@tib.eu}.  

\subsection{Code architecture}\label{4.2}

The main \pkg{dtreg} features are shown in Figure~\ref{fig_2}. First, the user selects a schema based on the data analysis method and creates a mutable instance of a schema-related class with the function \code{load_datatype()}. Next, the user populates the instance with relevant information about data analysis (method, data, and results). Finally, the user converts the instance into a machine-readable format with the function \code{to_jsonld()}. 

\begin{figure}
    \centering
    \includegraphics[width=0.6\linewidth]{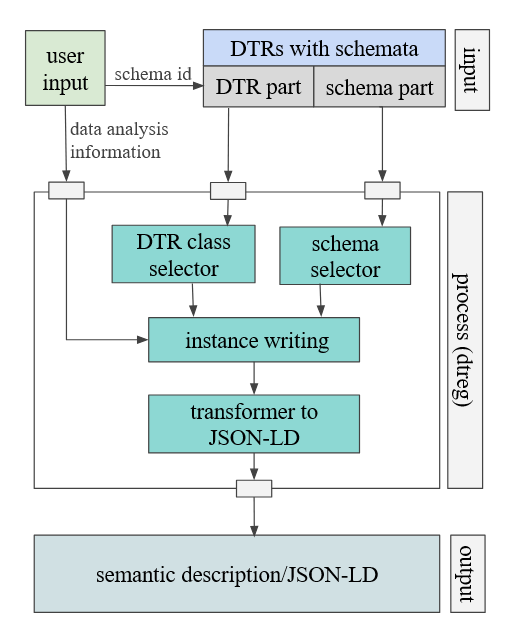}
    \caption{Information flow in the dtreg package using input-process-output (IPO) model.}
    \label{fig_3}
\end{figure}

Figure~\ref{fig_3} shows the information flow in more detail and gives insight into the code architecture. It specifies the user input, the process (the \pkg{dtreg} package functionality), and the output (the resulting machine-readable data).
\begin{figure}
    \centering
    \includegraphics[width=0.9\linewidth]{Fig2.png}
    \caption{The relationship between the user input and the dtreg functionality. The user selects the schema based on the data analysis method. The entirety of data analysis information (the method, the data, and the test results) is used to populate the instance.}
    \label{fig_2}
\end{figure}

When the user provides a schema URL, internal functions take its DTR prefix to select a DTR and the schema-id suffix to obtain the schema information; in Figure~\ref{fig_3}, these are “DTR class selector” and “schema selector”, respectively. The schema information is obtained from the static files, which store all ePIC data analysis schemata and their related sub schemata for fast and offline retrieval. When the schemata are updated, we make a new release of \pkg{dtreg}; therefore, we recommend always using the latest version of the package. If a schema is not in static files, \pkg{dtreg} makes an API request to the respective DTR (in Figure~\ref{fig_3}, “DTRs with schemata”).

The result of the \code{load_datatype()} function is a mutable instance of a schema-related class. In Python, we use Protocols \citep{typing_lib}, and in R, we use R6 classes \citep{r6}. To be more specific, the function returns a set of instances related to the hierarchy of schemata described here in Section 2. These instances are retrieved as a dictionary in Python (with SimpleNamespace for syntactic sugar), or as a named list in R. The names of these schemata can be checked with the \code{keys()} method in Python or \code{names()} in R. When the user populates the main instance (e.g., \code{group_comparison}) with data analysis information, its connected instances (\code{software_method}, \code{data_item}, etc.) can be easily included via the corresponding fields, as we show in the next section. Available fields for any schema can be checked via the \code{prop_list} attribute in Python or the \code{show_fields()} helper function in R. 

The finalized instance is converted into JSON-LD with the \code{to_jsonld()} function. Internally, the function handles different types of input provided by the user, enriching it with URI context and data type information required for mapping the data in JSON-LD. Finally, the Python or R object is serialized as a JSON string, for which we use the in-built module \pkg{json} in Python \citep{json_lib}, or the package \pkg{jsonlite} in R \citep{ooms2014jsonlite}. Users can save this result as a file, which they can upload to a repository or submit as supplementary materials to their paper.  

\subsection{Use case}\label{4.3}
For illustration purposes, let us assume that a researcher conducted a t-test comparing petal length of setosa and virginica species from Iris Data (Fisher 1936). The test code and other details can be found on our help page at \url{https://knowledgeloom.tib.eu/pages/help}. The results of the test include $t$~statistics, degrees of freedom, and the $p$~value written as a data frame (\code{df_results}). 

To report the results in a machine-readable format, \pkg{dtreg} should be installed from PyPi, e.g. using \code{pip} or \code{project.toml}. The researcher selects the schema and gets the URL from the help page to use as an argument in the \code{load_datatype()} function. The following loads the \code{group_comparison} schema. 
\begin{minted}
[
frame=single,
framesep=1mm,
fontsize=\footnotesize
]
{python}
from dtreg.load_datatype import load_datatype
dt_gc = load_datatype("https://doi.org/21.T11969/b9335ce2c99ed87735a6")
\end{minted}
Then, to populate the group comparison instance, the researcher describes: (i) the software method, (ii) the input data (here, as \code{data_url}); and (iii) the test results as a data frame (\code{df_results}). For the sake of simplicity, versions of Python and \code{scipy} are hardcoded here; in reality, they are obtained with \code{sys.version_info} and \code{importlib.metadata.version}, respectively. 
\begin{minted}
[
frame=single,
framesep=1mm,
fontsize=\footnotesize
]
{python}
instance_gc = dt_gc.group_comparison(
    label = "t-test Iris petal length setosa vs virginica",
    executes = dt_gc.software_method(
        label = "ttest_ind",
        is_implemented_by = "ttest_ind(setosa, virginica, equal_var = False)",
        part_of = dt_gc.software_library(label = "scipy",
                                         version_info = "1.15.1",
                                         part_of = dt_gc.software(label = "Python",
                                                                  version_info = "3.12.5"))),
    targets = dt_gc.component(label = "petal length (cm)"),
    has_input = dt_gc.data_item(label = "iris", source_url = "data_url"),
    has_output = dt_gc.data_item(source_table = df_results)
)
\end{minted}
The instance is mutable; we can change, for instance, the input data label to be more descriptive.
\begin{minted}
[
frame=single,
framesep=1mm,
fontsize=\footnotesize
]
{python}
instance_gc.has_input.label = "Iris petal length setosa virginica"
\end{minted}
Now, the \code{data_analysis} schema should be loaded. The data analysis instance contains all procedures conducted in the process of data analysis, in our case only the t-test, and the reference to the code (\code{code_url}).
\begin{minted}
[
frame=single,
framesep=1mm,
fontsize=\footnotesize
]
{python}
dt_da = load_datatype("https://doi.org/21.T11969/feeb33ad3e4440682a4d")
instance_da = dt_da.data_analysis(is_implemented_by = "code_url",
                                  has_part = instance_gc)
\end{minted}
A machine-readable representation of the data analysis instance in JSON-LD format is produced by calling the \code{to_jsonld()} function.
\begin{minted}
[
frame=single,
framesep=1mm,
fontsize=\footnotesize
]
{python}
from dtreg.to_jsonld import to_jsonld
ttest_json = to_jsonld(instance_da)
\end{minted}
The same procedure in R is given next without further comments, as it is similar to what is described above for Python. Similarly to the example above, changing the input label is not a part of the usual procedure, but merely an illustration that the instance is mutable. 
\begin{minted}
[
frame=single,
framesep=1mm,
fontsize=\footnotesize
]
{r}
library("dtreg")
dt_gc <- dtreg::load_datatype("https://doi.org/21.T11969/b9335ce2c99ed87735a6")
instance_gc <- dt_gc$group_comparison(
  label = "t-test Iris petal length setosa vs virginica",
  executes = dt_gc$software_method(
    label = "t.test",
    is_implemented_by = "stats::t.test(setosa, virginica)",
    part_of = dt_gc$software_library(
      label = "stats",
      version_info = "4.3.1",
      part_of = dt_gc$software(label = "R",
                               version_info = "4.3.1"))),
  targets = dt_gc$component(label = "Petal.Length"),
  has_input = dt_gc$data_item(label = "iris",
                              source_url = "data_url"),
  has_output = dt_gc$data_item(source_table = df_results))
instance_gc$has_input$label = "Iris petal length setosa virginica"
dt_da <- dtreg::load_datatype("https://doi.org/21.T11969/feeb33ad3e4440682a4d")
instance_da = dt_da$data_analysis(is_implemented_by = "code_url",
                                  has_part = instance_gc)
ttest_json <- dtreg::to_jsonld(instance_da)
\end{minted}
The resulting machine-readable representation is enriched with semantic context, such as data typing. It can be easily written as a file, which can be submitted to the TIB Knowledge Loom at \url{https://knowledgeloom.tib.eu/pages/submit} and shared publicly. This facilitates transparent reporting of research findings in a machine-readable format.

Research papers that have already been described in a machine-readable format with dtreg and presented for researchers in a human-readable way can be accessed at \url{https://knowledgeloom.tib.eu/}. A presentation of the use case discussed above, a t-test on Iris Data in Python, can be found under the title \textit{Analysis of difference for selected characteristics of Iris species}. We created this entry for illustration purposes to show that the reader can easily get information about the method, the data, and the test results, as well as download the data and the code.

\section{Limitations and future work}

Although the \pkg{dtreg} package in the current version is stable, new releases might be required to accommodate schemata modifications, therefore we recommend that users install the latest version of the package. For instance, we might modify the categorization of data analysis methods or increase categorization granularity; improve the semantic interoperability of the data via ontological terminology; or add some methods, such as Bayesian statistics, time series, etc., that are yet to be covered by the approach. A new release is also possible if we add another DTR to the two that are currently supported. 

As can be seen from section \ref{4.3}, converting research findings into a machine-readable format with \pkg{dtreg} is easy; however, it is important to populate the instance with sufficiently detailed data analysis information. Therefore, we are developing a new package based on \pkg{dtreg} with wrapper functions that make the process of populating the instance even easier for researchers\footnote{The package \pkg{mrap} is already released in PyPi \url{https://pypi.org/project/mrap/} and CRAN \url{https://cran.r-project.org/package=mrap}.}. The user is spared the effort of writing the information that can be taken from the computing environment (the software version etc.) but able to change the resulting object to keep full control over the process and resulting data.

Currently, researchers might use \pkg{dtreg} with data types that they construct in one of the supported DTRs, and those conducting quantitative data analysis can apply data analytic schemata included in the package. An extension of applicability is possible by means of including other DTRs, in addition to the ePIC and ORKG DTRs, which the package currently interacts with. Also, from the perspective of data management, \pkg{dtreg} can be smoothly integrated in a workflow management system \citep{wilkinson2025applying}.

\section{Concluding remarks}
In this paper, we introduced the \pkg{dtreg} package in Python and R. The package supports writing research findings in a machine-readable format. For this purpose, information about data analysis is structured with registered and persistently identified data types, aka schemata, and the resulting object is converted into JSON-LD. We explained the code architecture, illustrated the \pkg{dtreg} functionality with a use case, showed the reuse potential of the package, and outlined the directions of future work. We suggest that the \pkg{dtreg} package can enhance the methodological repertoire of scientists from any domains, who aim to adhere to the FAIR principles, write their research findings in a machine-readable format, and report them transparently.   

\section*{Acknowledgments}
The authors would like to express their appreciation to Lars Vogt for his invaluable help with schemata, which advanced us towards semantic interoperability of the data, and to Lauren Snyder, who provided us with user-friendly formulations for the package documentation. 

\section*{Funding statement}

This work was supported by the German Research Foundation (DFG) project NFDI4DS (PN: 460234259).

\section*{Competing interests}

The authors have no competing interests.

\bibliography{refs}

\end{document}